\documentclass[a4paper,preprint,10pt,3p,twocolumn]{elsarticle}
\usepackage[english]{babel}
\usepackage{graphicx}
\usepackage{color}
\usepackage{amssymb}
\usepackage{amsmath}
\usepackage{enumitem}
\usepackage{array}
\usepackage{booktabs}

\newcommand{\be}{\begin{equation}}
\newcommand{\ee}{\end{equation}}
\newcommand{\ba}{\begin{eqnarray}}
\newcommand{\ea}{\end{eqnarray}}

\begin{document}

\title{Equilibration and freeze-out of an expanding gas in a transport approach in a Friedmann-Robertson-Walker metric}

\author{J.~Tindall$^{1,2}$, J.M. Torres-Rincon$^{2}$, J.B. Rose$^{2,3}$, H. Petersen$^{2,3,4}$}

\address{$^1$Department of Physics, University of Bath, Claverton Down, Bath, United Kingdom, \\
$^2$Frankfurt Institute for Advanced Studies, Ruth-Moufang-Str. 1, 60438, Frankfurt am Main, Germany, \\
$^3$Institute for Theoretical Physics, Goethe University, Max-von-Laue-Strasse 1, 60438 Frankfurt am Main, Germany, \\
$^4$GSI Helmholtzzentrum f\"ur Schwerionenforschung, Planckstr. 1, 64291 Darmstadt, Germany}

\begin{abstract}

Motivated by a recent finding of an exact solution of the relativistic Boltzmann equation in a Friedmann-Robertson-Walker spacetime, we implement this metric 
into the newly developed transport approach Simulating Many Accelerated Strongly-interacting Hadrons (SMASH). We study the numerical solution of the transport equation and compare it to this exact solution for massless particles. 
We also compare a different initial condition, for which the transport equation can be independently solved numerically. Very nice agreement is observed in both cases. Having passed these checks for the SMASH code, 
we study a gas of massive particles within the same spacetime, where the particle decoupling is forced by the Hubble expansion. In this simple scenario we present an analysis 
of the freeze-out times, as function of the masses and cross sections of the particles. The results might be of interest for their potential application to relativistic heavy-ion collisions, for the
characterization of the freeze-out process in terms of hadron properties.
\end{abstract}


\maketitle

\section{Introduction}

  Kinetic theory~\cite{lifschitz1983physical} has been widely used to study the nonequilibrium evolution of fluids and plasmas, not only for ordinary substances but also in the relativistic domain~\cite{de1980relativistic}. 
For sufficiently dilute systems, the Boltzmann equation (BE) describes how the one-particle distribution function $f(t,{\bf x},{\bf k})$ relaxes towards equilibrium. Under a general spacetime metric, this equation reads~\cite{Cercignani}
\begin{equation} \label{eq:BE}
   k^{\mu} \frac{\partial f (t,{\bf x},{\bf k})}{\partial x^\mu} + \Gamma_{\lambda \mu}^{i} \ k^{\lambda} k^{\mu} \frac{\partial f (t,{\bf x},{\bf k})}{\partial k^{i}} = C[f] \ ,
\end{equation} 
where $\Gamma_{\lambda \mu}^i$ are the Christoffel symbols and $C[f]$ represents the (nonlinear) collision integral~\cite{de1980relativistic}. 

  A non-trivial solution of this equation (aside from the equilibrium distribution) is extremely hard to obtain in this general case. Nevertheless, several approximations can be used to simplify the nonlinear structure
of $C[f]$. One of the simplest methods is the relaxation time approximation (RTA)~\cite{Bhatnagar:1954zz}, which provides a linearized collision term. In addition, perturbative solutions based on the existence of
some small parameter (like the Knudsen number) are also
possible e.g. the Chapman-Enskog expansion~\cite{chapman1952mathematical,Torres-Rincon:2012sda}. On the other hand, the BE can also be addressed by pure numerical techniques, known as molecular dynamics simulations
or Boltzmann-Uehling-Uhlenbeck (BUU) transport. 
For relativistic heavy-ion collisions (RHICs), where a mixture of relativistic particles is subjected to mutual interactions and mean-field potentials, the system of coupled BEs can be solved by Monte Carlo methods, as
in ~\cite{Bass:1998ca,Lin:2004en,Cassing:2009vt,Buss:2011mx,Nara:2016phs,Weil:2016zrk}.
These numerical approaches---more suitable for these complicated systems---also introduce systematic uncertainties, originating from algorithmic approximations and truncations. 
For this reason, the finding of exact non-trivial solutions of Eq.~(\ref{eq:BE}), at least in particular scenarios, is important to test different methods and approximations, either semianalytical or purely numerical.

  In RHICs the dynamics and geometry of the created fireball provide certain degrees of symmetry, from which simplified models have been proposed~\cite{Csernai:1994xw}. For some of them, exact solutions
of the BE have been found under the RTA. For example, in the Bjorken model~\cite{Bjorken:1982qr} (describing a boost-invariant longitudinal expansion) a semianalytic result has already been calculated by Baym~\cite{Baym:1984np}.
An exact solution in a Gubser expansion (allowing an additional expansion in the transverse plane) has been obtained in~\cite{Denicol:2014xca}. An exact solution in a 3D conformal expanding medium, or Hubble flow, 
was recently found in~\cite{Noronha:2015jia,Hatta:2015kia} under the RTA.

  This last scenario represents an interesting case with applications to RHICs but also in a cosmological context to describe an expanding universe~\cite{Baumann,Kolb,Bernstein}. An exact analytical solution of the BE for 
an expanding medium has recently been presented in~\cite{Bazow:2015dha,Bazow:2016oky}. This solution is valid for massless particles, interacting in a flat universe under a Friedmann-Robertson-Walker (FRW) metric. The 
particle-particle interactions are assumed to follow constant total cross sections, but the full nonlinear structure of the collision operator is kept, i.e. no RTA is assumed. The exact solution obtained
in~\cite{Bazow:2015dha,Bazow:2016oky} was used to test a linear approximation of the BE, and approximate solutions based on the RTA.
  
  In this paper we compare a numerical solution of Eq.~(\ref{eq:BE}) with this exact result. We employ the new hadronic transport approach SMASH
(Simulating Many Accelerated Strongly-interacting Hadrons)~\cite{Weil:2016zrk}. It is used to simulate hot and dense strongly interacting matter with the goal of exploring the quark gluon plasma phase diagram by
performing comparisons to experimental data from heavy ion experiments at different accelerators such as the Large Hadron Collider (LHC), the Relativistic Heavy Ion Collider and the SIS-18 at the
GSI Helmholtzzentrum f\"ur Schwerionenforschung.
SMASH constitutes an effective numerical solution of the equations of motion associated with Eq.~(\ref{eq:BE}) using a geometrical collision criterion as in UrQMD\footnote{Ultra-relativistic Quantum Molecular Dynamics.}
under a Minkowski metric, i.e. $ds^2 = dt^2 -dx^2 -dy^2 -dz^2$. We adapt the SMASH dynamics to work on an expanding homogeneous, isotropic gas of massless particles with a FRW metric $ds^2 = dt^2 -a^2(t) (dx^2 +dy^2 +dz^2)$
\footnote{We only consider the case without spatial curvature $K$.}. In this case the BE is reduced to~\cite{Bernstein,Cercignani,Bazow:2016oky}
\begin{equation} \label{eq:BEsimp}
     k^{\mu} u_\mu u_\nu \partial^\nu f (t,k) = C_{gain}[f]-C_{loss}[f] \ ,
\end{equation}
where the gain and loss terms present their full nonlinear structure.

  Our first goal is to present a non-trivial test of the SMASH code in an expanding geometry by comparing our outcome to the exact solution given in~\cite{Bazow:2015dha,Bazow:2016oky}. 
We also check our results against the numerical solution of Eq.~(\ref{eq:BEsimp}) for a different initial condition given in \cite{Bazow:2016oky}. Then, we exploit the flexibility of SMASH
to solve the transport equation in an expanding system of massive particles, generating a dynamical freeze-out (or decoupling) due to the Hubble expansion. 
This opens up the possibility to study more realistic systems of interest in cosmological scenarios, or in RHICs.

In Sec.~\ref{sec:solution} we present the SMASH solution to the Boltzmann equation for massless particles using several initial conditions, in particular the one for which an exact analytical solution is known.
In Sec.~\ref{sec:freezeout} we introduce a toy model of freeze-out for relativistic particles when the Hubble rate exceeds the interaction rate. We discuss how the freeze-out time can be extracted from the final
spectrum of particles. Finally, we present our conclusions and outlook in Sec.~\ref{sec:summary}.

%%%%%%%%%%%%%%%%%%%%%%%%%%%%%%%%%%%%%%%%%%%%%%%%%%%%%%%%%%%%%%%%%%%%%%%%%%%%%%%%%%%%%%
%%%%%%%%%%%%%%%%%%%%%%%%%%%%%%%%%%%%%%%%%%%%%%%%%%%%%%%%%%%%%%%%%%%%%%%%%%%%%%%%%%%%%%

\section{\label{sec:solution} SMASH solution of the Boltzmann equation under a FRW spacetime}

The authors of Ref.~\cite{Bazow:2015dha,Bazow:2016oky} have calculated an exact solution of the Boltzmann equation~(\ref{eq:BEsimp}) for an infinite gas of massless particles with constant
elastic cross-section. This is a very particular system, with symmetry properties that help to simplify the transport equation. This scenario is physically motivated by the
expansion of the universe in the radiation-dominated era~\cite{Baumann}.

SMASH is a recently developed transport approach used to describe the hadronic stage of heavy-ion collisions at low and intermediate energies with applications from GSI to LHC physics. In particular, one can use SMASH to simulate a gas 
of massless particles with constant elastic cross-section $\sigma$. In this work we use a spherical volume filled with $N$ particles (this number remains constant due to the absence of number-changing processes).
The particles are initialized with an isotropic, homogeneous spatial distribution according to a given initial condition. 

Whilst it appears a formidable task to adapt SMASH to a general spacetime, the FRW metric we wish to implement is fairly simple. As SMASH operates in physical phase-space variables,
we will always work with the physical 3-momenta $k=k_{phys}$, as opposed to the approach in~\cite{Bazow:2015dha,Bazow:2016oky}, which uses the covariant momenta\footnote{The distinction is important 
in nonorthonormal metrics~\cite{Kolb}. In our equations we will always trade the modulus of the covariant momentum---whose magnitude is not 
modified by the expansion of the system---by the physical one.}.  
The equations of motion of the particles reflect the physical expansion of the universe. The velocity measured by a comoving observer to the expanding spacetime (sometimes called peculiar velocity)
is combined with the Hubble flow:
$v_{Hubble}^i = H(t) x^i$, where $H(t) \equiv \frac{\dot{a}(t)}{a(t)}$ is the Hubble parameter. The physical momentum of the particle suffers a redshift and scales as $1/a(t)$.

Particle collisions are not affected by the Hubble expansion, because the characteristic collision time is always much smaller than $H^{-1}(t)$, so during the collision the particles do not feel the expansion 
of the universe.

For a gas of massless particles (radiation) the cosmic scale factor is fixed by the Friedmann equation to be of the form $a(t) \sim t^\frac{1}{2}$~\cite{Baumann,Bernstein,Kolb}.
Following~\cite{Bazow:2015dha,Bazow:2016oky} we adopt the solution $a(t) = \sqrt{1 + \frac{b_{r}}{l_{0}}t}$, where $l_{0}=1/(\sigma n_0)$ is the mean-free path at time $t = 0$ ($\sigma$ denotes the cross section and $n_0$ the initial 
particle density) and $b_{r}$ is a parameter
which contains the density fraction of radiation in the universe and the Hubble parameter itself at $t=0$.

We initialize the particles in a far-from-equilibrium configuration, according to the momentum distribution in~\cite{Bazow:2015dha,Bazow:2016oky} 
\be \label{eq:icexact} f(t=0,k)= \frac{256}{243} \frac{ka}{T_0} \lambda \exp \left( - \frac{4ka}{3T_0} \right) \ , \ee
where $\lambda=\exp(\mu_0/T_0)$ is the fugacity of the system, and $T_0$ a parameter, which can be thought of as an initial temperature of
the system\footnote{Out of equilibrium, the expressions of the particle and energy densities formally coincide with that for an ideal gas, with $T_0$ playing the role of the temperature.}.

The analytical solution for this initial condition was determined to be
\begin{equation} \label{eq:exact}
f (t,k) = \lambda \frac{e^{-\frac{k a}{\kappa T_{0}}}}{\kappa^4 (\tau)} [4\kappa - 3 + \frac{k a}{\kappa (\tau) T_{0}} (1 - \kappa (\tau))] \ , 
\end{equation}
where $\kappa (\tau) = 1 - \exp \left(-\frac{\tau}{6} \right)/4$, and the transformed time variable $\tau =\int_{\hat{t_{0}}}^{\hat{t}} \frac{1}{a^3 (\hat{t}')} d\hat{t}'$ with $\hat{t} = t/l_{0}$. For the particular form 
of the $a(t)$ used, we have
\begin{equation}
\tau = \frac{2}{b_{r}}\left[ 1 - \left( 1 + \frac{b_{r}}{l_{0}}t \right)^{-1/2} \right] \ .
\end{equation}

The distribution function is normalized such that 
\begin{equation}
N= \int_{V} \int f (t,k) \frac{d^3k}{(2\pi)^3} d^3 {\bf x} \ , 
\end{equation}
where $V$ is the volume of the sphere, and $d^3k = dkk^2 d\Omega_{k}$.

We now present the results of SMASH for the solution of the Boltzmann equation for different values of the physical time. Notice that we show these results in terms of $k$ multiplied by the scale factor.

\begin{figure}[t]
 \includegraphics[scale=0.16]{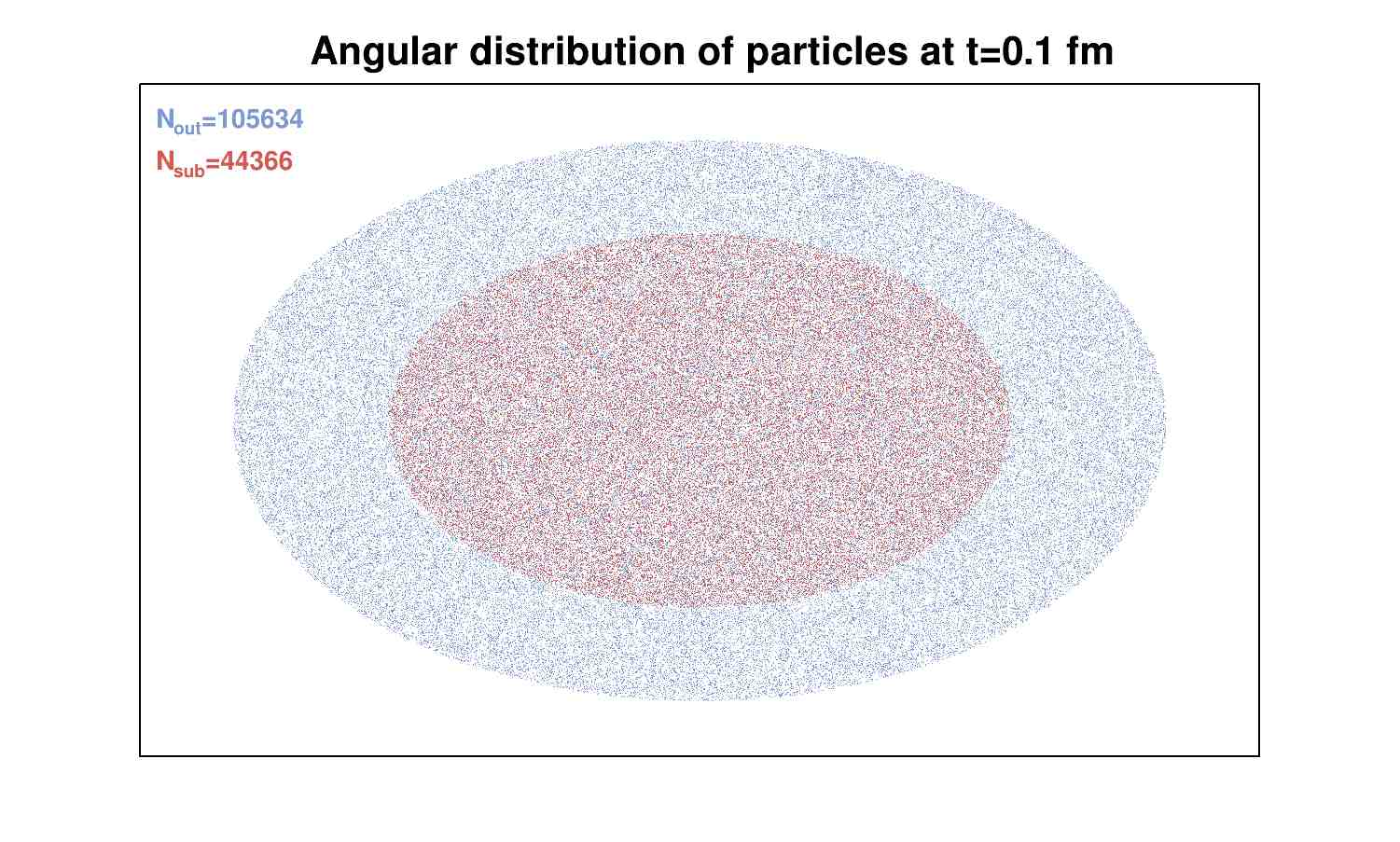}
\caption{\label{fig:Mollweide} Mollweide projection of the angular distribution of particles at $t=0.1$ fm. We show all particles contained in the sub-sphere $r<r_{sub}$ in red, and the particle outside this volume ($r>r_{sub}$)
in blue (and re-scaled by a factor 1.5 for clarity). We use a single event $N_{ev}=1$.}
\end{figure}

In our simulations we use a constant particle number $N=1.5 \times 10^5$ contained in a spherical volume with initial radius $r_0=50$ fm. The ratio $b_{r}/l_{0}$ is fixed to 0.1. The initial temperature of the system is set to $T_{0} = 0.2$ GeV. With this set of parameters, the fugacity takes a value of $\lambda_0 \simeq 2.7$, which
should be constant in the evolution, cf. Eq.~(\ref{eq:exact}).  To avoid unwanted boundary effects and ensure homogeneity we always work with particles contained in a sub-sphere of radius $r_{sub}(t) = a(t) r_0 /1.5$.
To check the spatial isotropy we plot in Fig.~\ref{fig:Mollweide} the Mollweide projection of the distribution of particles for $t=0.1$ fm. The Mollweide projection reflects the 2D angular distribution 
of particles (neglecting their distance to the origin) into a 2D ellipsoidal plot. It is widely used in cartography and cosmology, but it can also be applied to heavy-ion physics~\cite{Llanes-Estrada:2016pso}\footnote{We thank Felipe Llanes-Estrada
for discussions on this topic.}.
In red we show the particles contained in the sub-sphere $r<r_{sub}$. In blue we depict the particles with $r>r_{sub}$, where we have re-scaled the Mollweide projection by the factor 1.5. Isotropy is manifest 
in the plot.

\begin{figure}[t]
\includegraphics[scale=0.36]{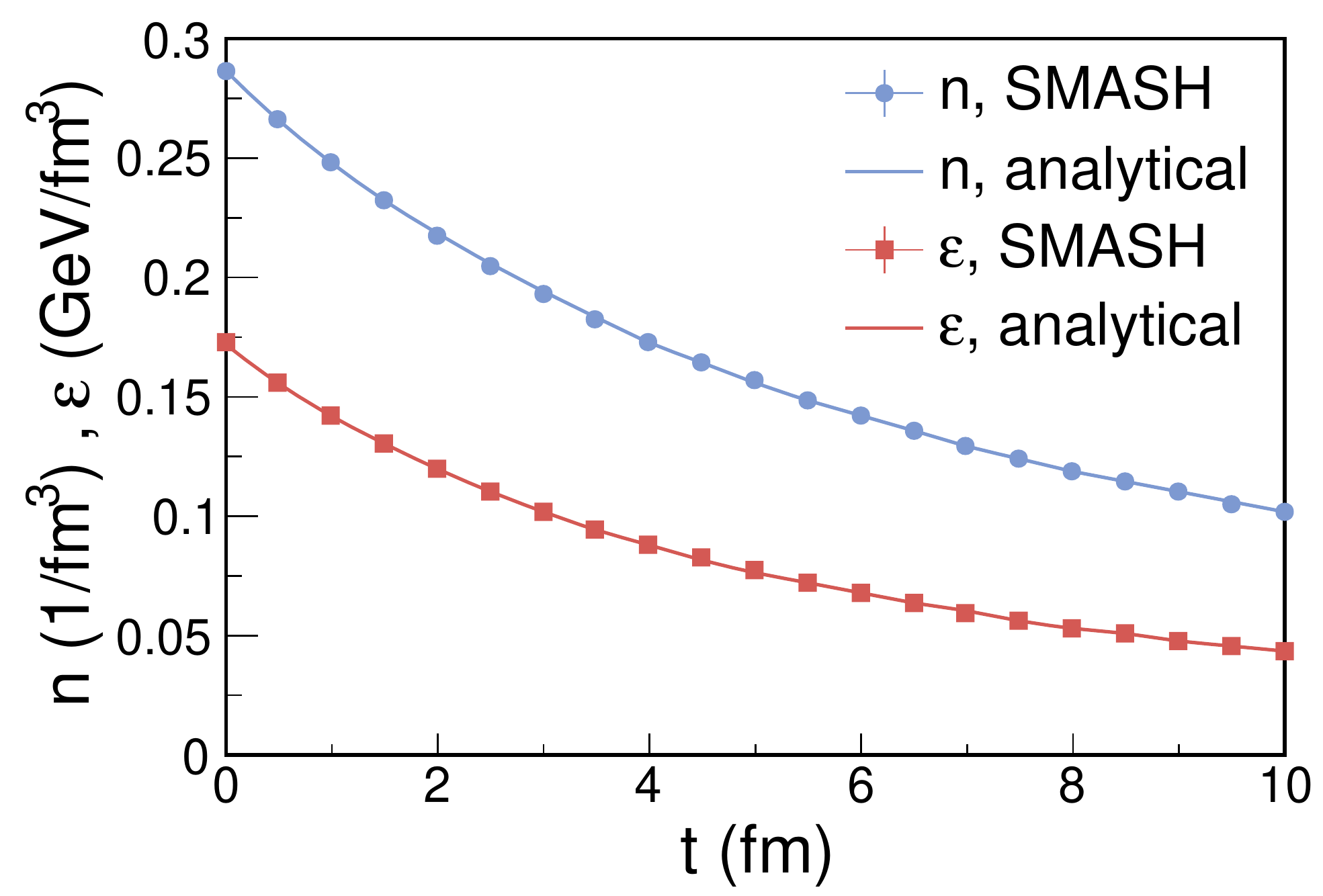}
\caption{\label{fig:densities} Time evolution of the particle and energy densities in SMASH (dots) for one event, compared to the analytical expression in Eqs.~(\ref{eq:densities}).}
\end{figure}

Before presenting our results for the distribution function we show the time evolution of the particle and energy densities in Fig.~\ref{fig:densities}. The theoretical expressions of the
(non-equilibrium) densities admit an analytic form~\cite{Bazow:2016oky},
\be \label{eq:densities} n = \frac{\lambda_0 T_0^3}{\pi^2 a^3} \ , \quad \varepsilon = \frac{3 \lambda_0 T_0^4}{\pi^2 a^4} \ . \ee
Our results are in very good accordance with these formulas, even for a single event.

\begin{figure}[htp]
\begin{center}
\includegraphics[scale=0.36]{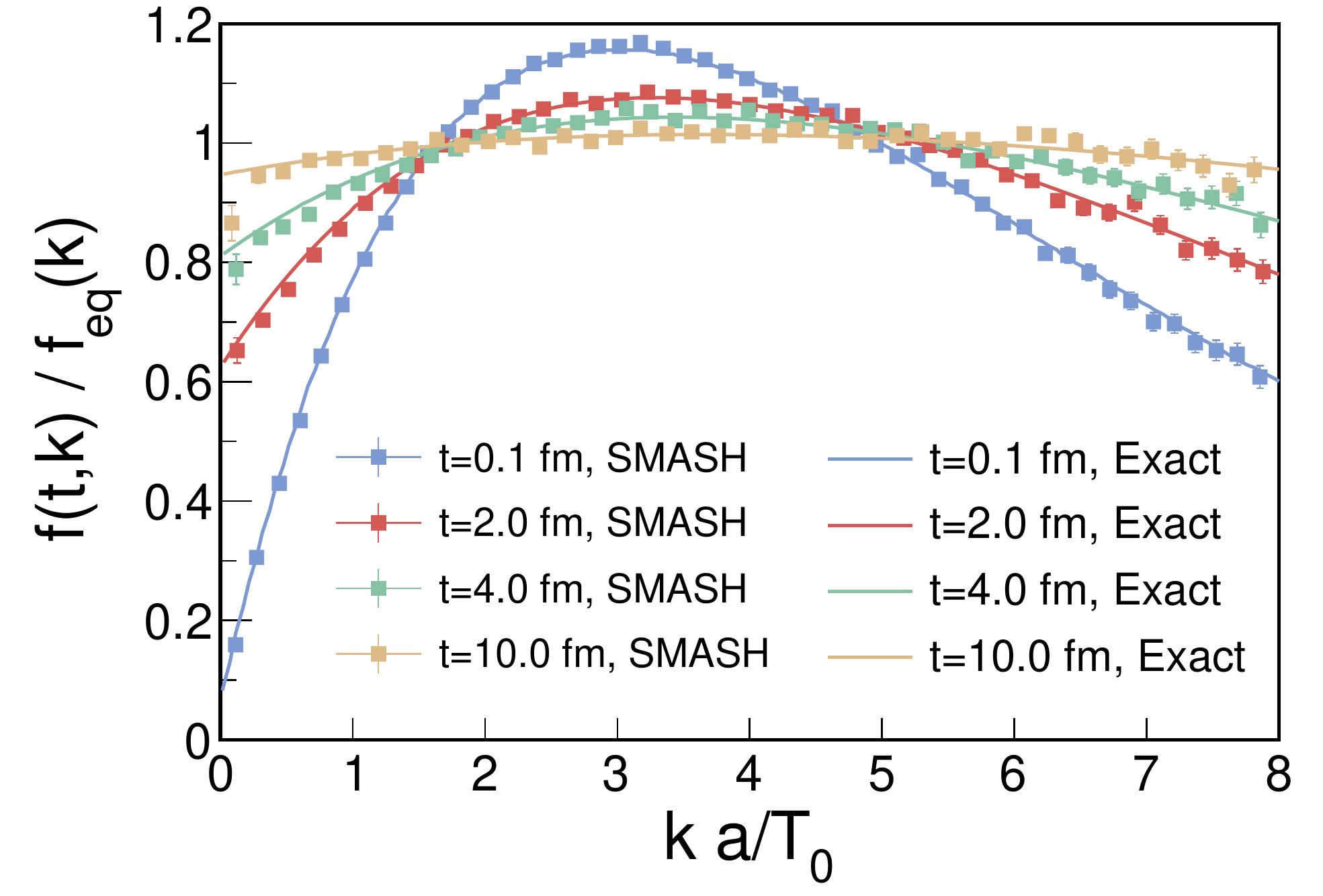}
\caption{\label{fig:exact} Ratio of the distribution function over the equilibrium Boltzmann distribution versus $\frac{ka}{T_{0}}$ for the initial condition~(\ref{eq:icexact}). The SMASH output (points) is compared to the
analytical solution of the Boltzmann equation (solid line) for different values of time $t=0.1,2,4,10$ fm.}
\end{center}
\end{figure}

In Fig.~\ref{fig:exact} we present the ratio $f (t,k)/f_{eq} (k)$, where $f_{eq} (k)= e^{-\frac{ka}{T_{0}}}$ is the (Boltzmann) equilibrium distribution
corresponding to $\lim_{\tau \to\infty} f(t,k)$\footnote{While the exact Boltzmann solution cannot be reached at $t\rightarrow \infty$ due to a finite $\tau$ horizon~\cite{Bazow:2016oky}, we still use it 
for normalization.}. We plot the results as a function of $\frac{ka}{T_{0}}$ for $t=0.1,2,4,10$ fm. 
We have used 20 events, and present up to the value $\frac{ka}{T_{0}} = 8.0$, where statistical errors start to become significant. The output from SMASH is compared to the analytical solution (\ref{eq:exact}) which uses the initial momentum distribution in Eq.~(\ref{eq:icexact}).
The results demonstrate a very good agreement between SMASH and the exact solution. Statistical errors due to the low number of particles in each bin explain the deviations between the analytical solution
and that of SMASH for $\frac{ka}{T_{0}} > 5$. For the lower momentum modes, $\frac{ka}{T_{0}} < 1$, the small deviations are a combination of statistical errors as well as the finite size of the bin-width which
has a noticeable effect in regions where $f(t,k)$ is changing rapidly with respect to $k$. 

\begin{figure}[htp]
\begin{center}
\includegraphics[scale=0.36]{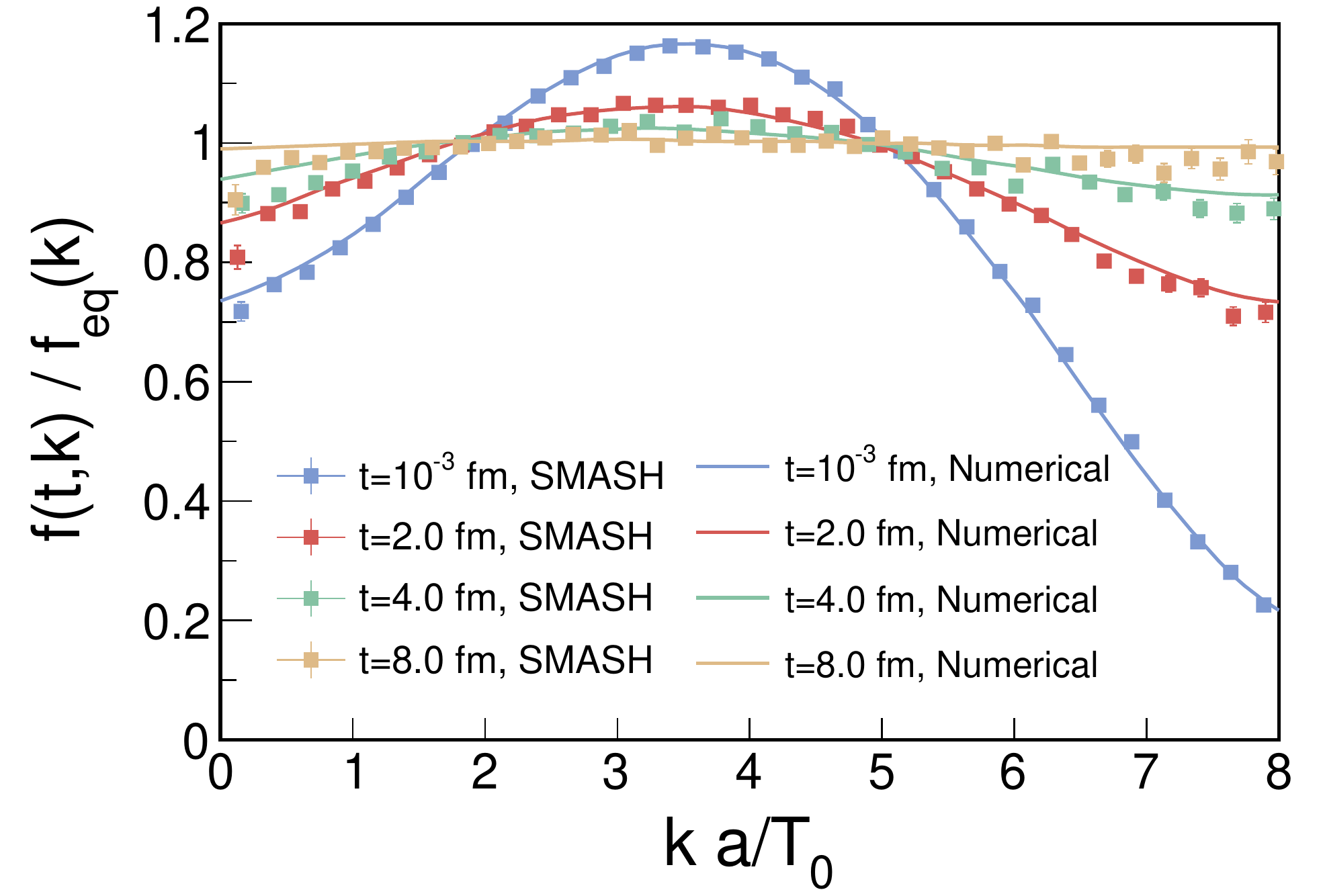}
\caption{\label{fig:2mode} Ratio of the distribution function over the equilibrium Boltzmann distribution versus $\frac{ka}{T_{0}}$ for the initial condition~(\ref{eq:ic2mode}). The SMASH result (points) is compared to the
numerical solution of the Boltzmann equation (solid line) for different values of time $t=0.001,2,4,8$ fm.}
\end{center}
\end{figure}

In Fig.~\ref{fig:2mode} a similar comparison is made with a different initial momentum distribution. We adapt the ``two-mode initial condition'' introduced in~\cite{Bazow:2016oky}:
\ba f(t = 0,k) &=&  \left[ 1 - \frac{1}{10} {\cal L}_{3}^{(2)} \left( \frac{ka}{T_{0}} \right) \right. \nonumber \\
&+& \left. \frac{1}{20} {\cal L}_{4}^{(2)} \left( \frac{ka}{T_{0}} \right) \right] \ e^{-\frac{ka}{T_{0}}} \ , \label{eq:ic2mode} \ea
where ${\cal L}_{n}^{(\beta)} (x) = \sum_{i = 0}^{n } (-1)^i \binom{n + \beta}{n - i} \frac{x^i}{i!}$ are Laguerre polynomials~\cite{de1980relativistic}. While no analytical solution
for this initial condition has been achieved, one can still numerically solve the BE for it. The latter, when written as a coupled system of equations for the moments of the distribution, can be easily 
solved by a finite difference method. From Fig.~\ref{fig:2mode} one can see that SMASH results and the numerical solution compare very well. Errors stemming from the binning procedure are more
significant here and cause the low-$k$ data points to noticeably deviate from the expected value. Overall this figure, together with Fig.~\ref{fig:exact}, can be considered a validation check for the SMASH approach to solving the relativistic
Boltzmann equation in a FRW metric.

%%%%%%%%%%%%%%%%%%%%%%%%%%%%%%%%%%%%%%%%%%%%%%%%%%%%%%%%%%%%%%%%%%%%%%%%%%%%%%%%%%%%%%
%%%%%%%%%%%%%%%%%%%%%%%%%%%%%%%%%%%%%%%%%%%%%%%%%%%%%%%%%%%%%%%%%%%%%%%%%%%%%%%%%%%%%%

\section{\label{sec:freezeout} Dynamical decoupling of a relativistic massive gas}

In this section we present a simple realization for the freeze-out of a particle; exploiting the Hubble expansion as a mechanism to dynamically decouple the particles at a certain time. The fundamental idea relies on
the fact that the scattering rate $\Gamma \simeq |v| n \sigma$~\footnote{$|v|$ is the relative velocity of the colliding particles} decreases in time due to the volume expansion, so that there might exist a time at which
this rate becomes smaller than the Hubble rate $H$ and the particles decouple.
This is the basic picture behind the description of the thermal history of the primitive universe~\cite{Baumann,Kolb,Bernstein}.

We implement a much simpler scenario, because we only consider a single species of relativistic particles interacting by a constant cross section.
The initial temperature is set to $T_0=0.4$ GeV and the mass of the particles to $m=5T_0$, so we deal with a relativistic gas of particles. One should note that a gas of ultrarelativistic or nonrelativistic particles is not 
useful in this study, as we will see that it is not possible to distinguish an equilibrium distribution evolving in time from a decoupled distribution with a momentum redshift~\cite{Baumann,Kolb}.

For relativistic particles, it is known that there exists no equilibrium distribution in the collisionless case (e.g. after decoupling)\footnote{More technically,
this statement is closely related to the nonexistence of a timelike Killing vector for a FRW metric with a nonconstant $a(t)$.}. Therefore, our toy model is based on the following scenario:
We start with an equilibrium distribution of massive particles. As long as $\Gamma \gg H$, collisions allow the system to maintain equilibrium despite of the Hubble expansion.
Once $\Gamma \sim H$, the number of collisions will not be enough to re-equilibrate the system, and the distribution function of particles freezes. For later times, when $\Gamma < H$ the 
distribution of particles is necessarily out of equilibrium due to the momentum redshift. 
To be able to see the decoupling within a reasonable amount of time we take the Hubble rate as a pure parameter of the model and use $H=\beta t$, with $\beta=0.01$ fm$^{-2}$. Notice that this is a rather arbitrary choice.
Our goal, however, is to establish a toy model for the freeze-out in RHICs, with a Hubble parameter encoding the physics of the expanding medium. For this exploratory study, we choose this parameter to illustrate
the effect of decoupling in a clear manner.

\begin{figure*}[htp]
\begin{center}
\includegraphics[scale=0.26]{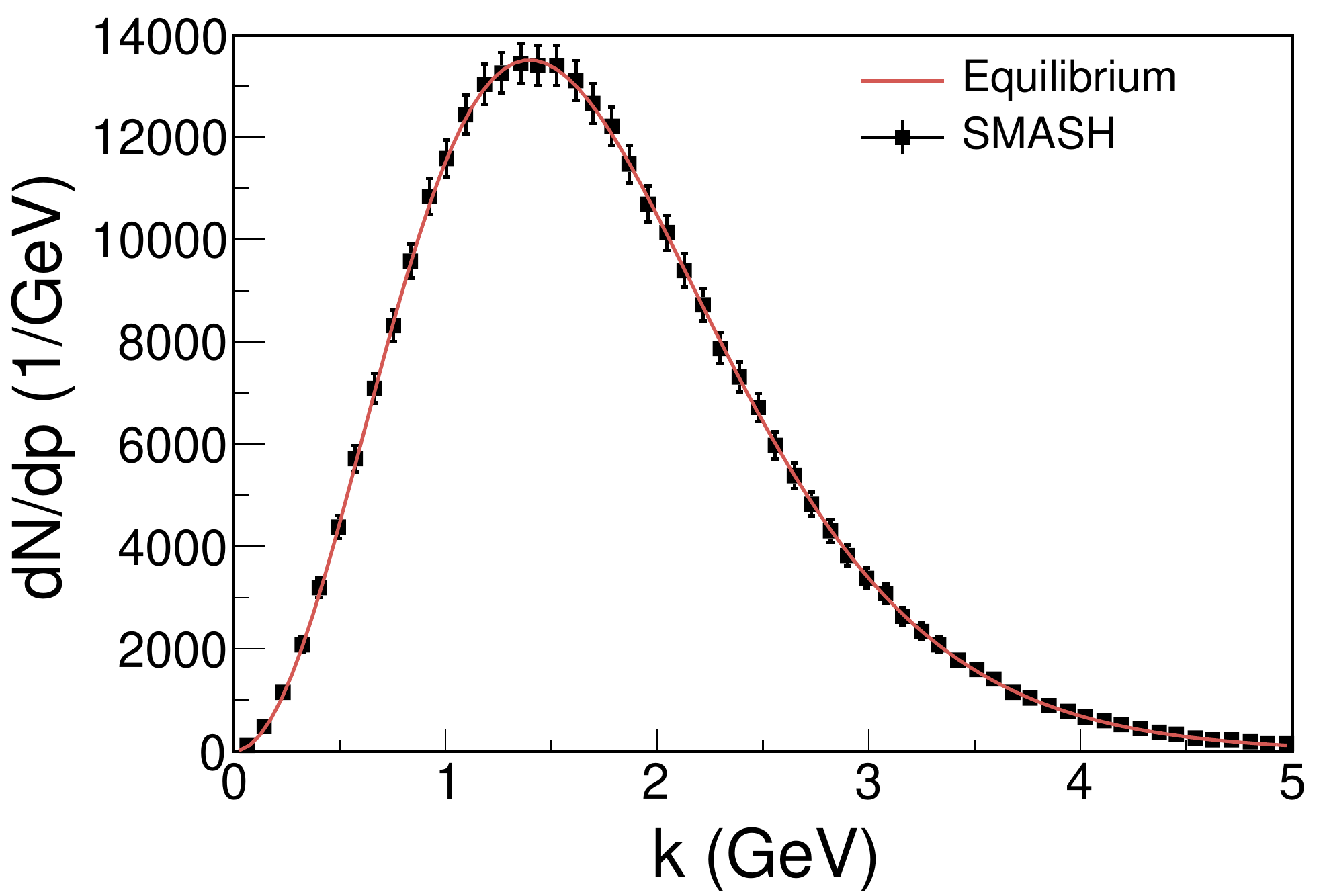}
\includegraphics[scale=0.26]{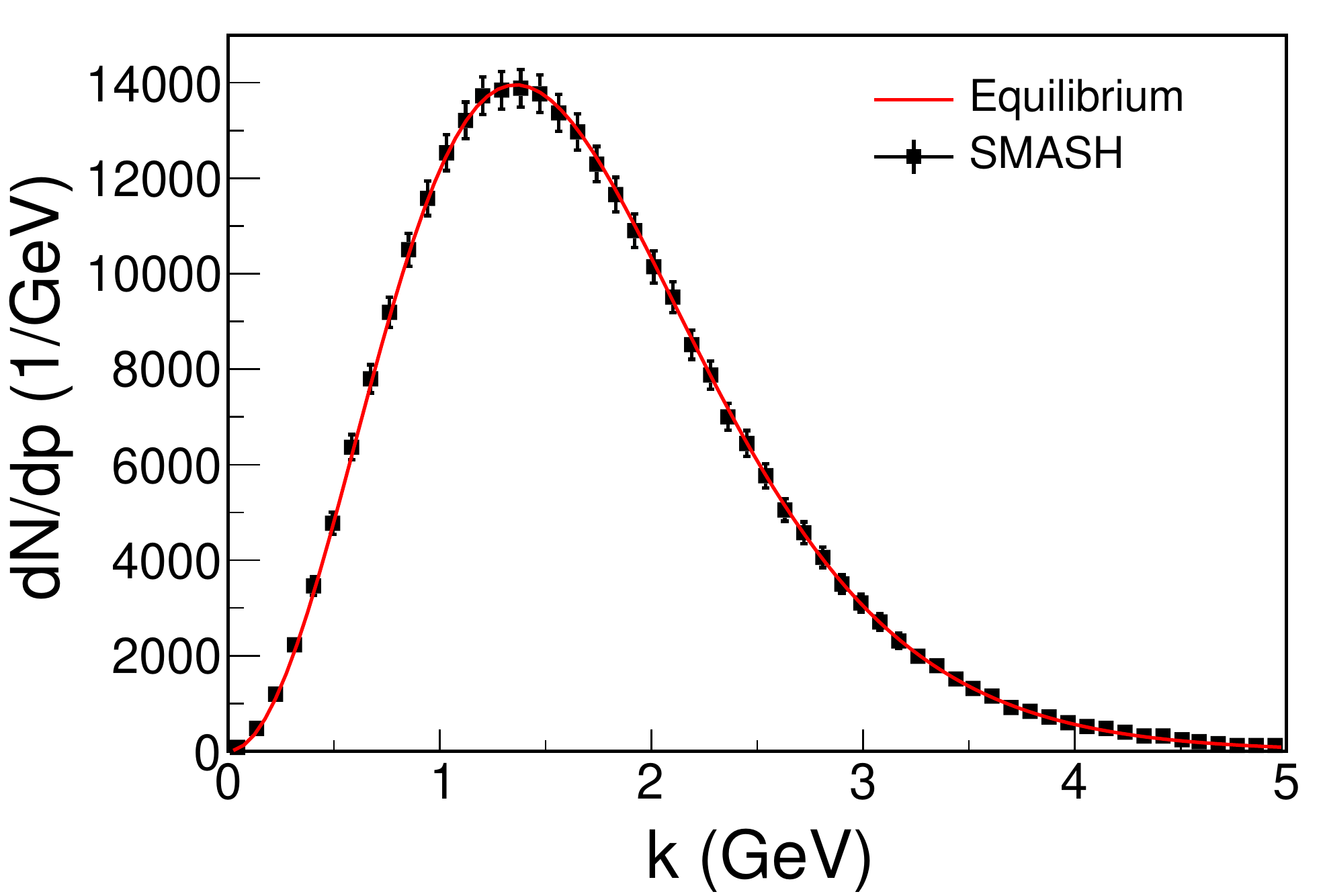}
\includegraphics[scale=0.26]{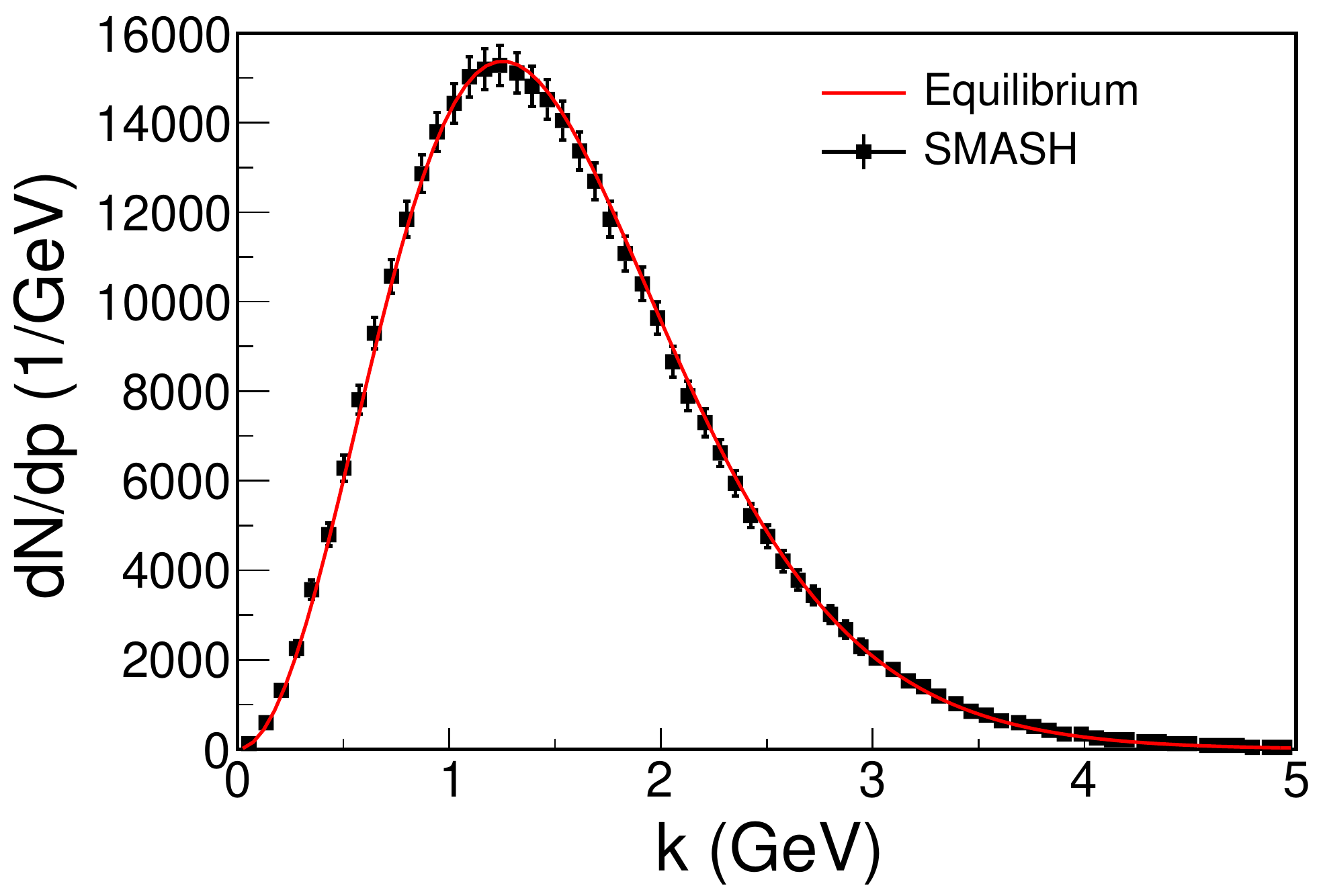}
\includegraphics[scale=0.26]{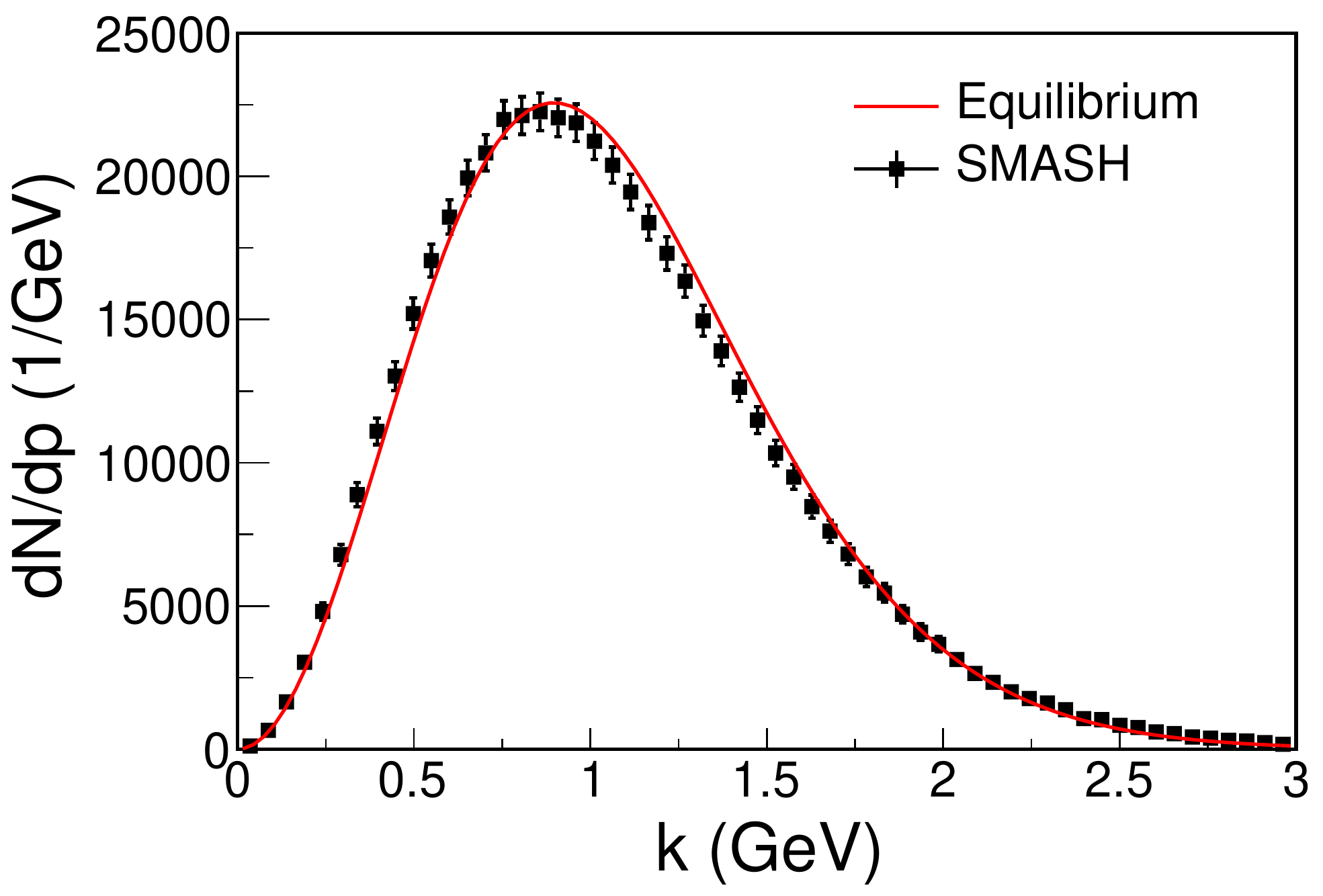}
\includegraphics[scale=0.26]{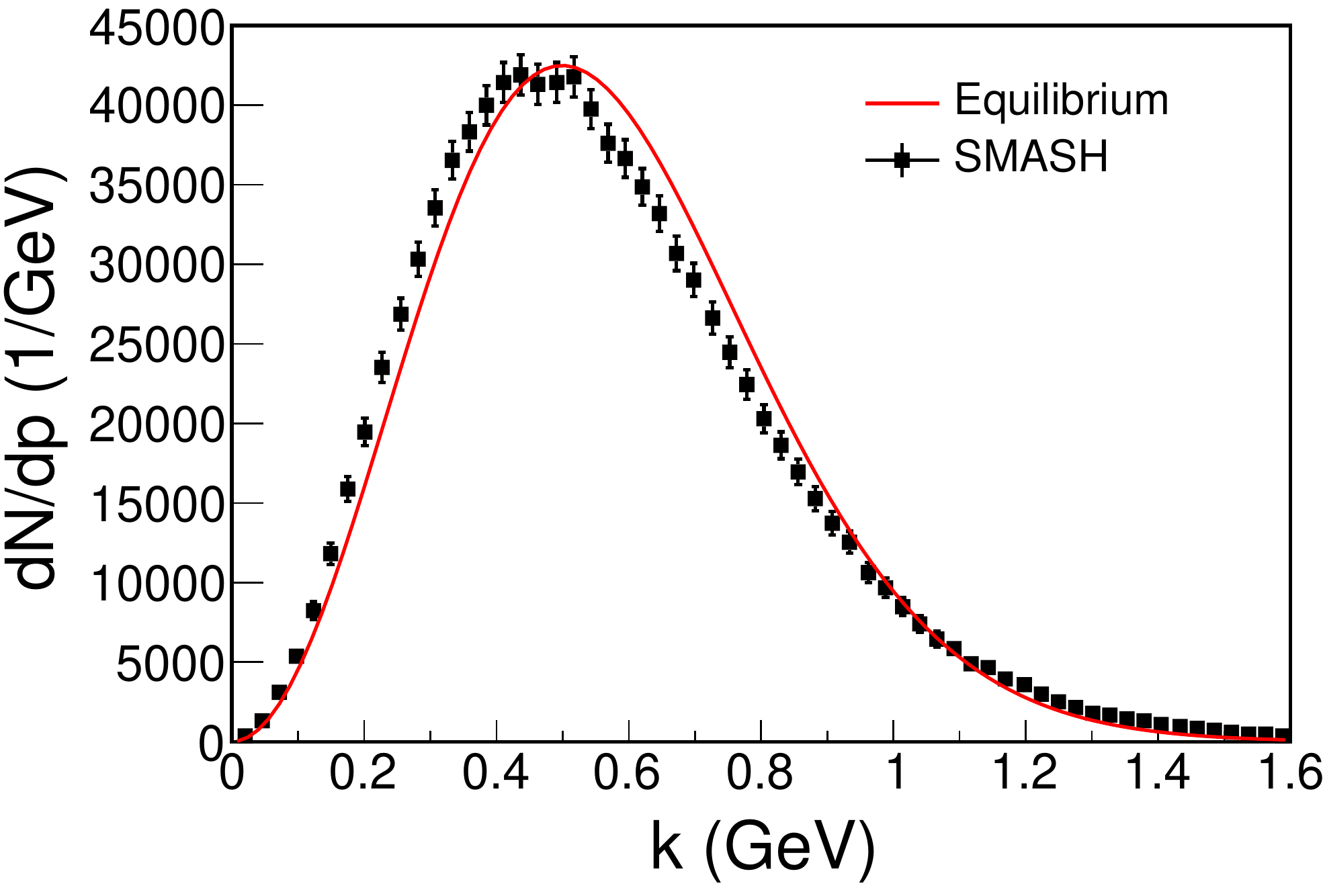}
\includegraphics[scale=0.26]{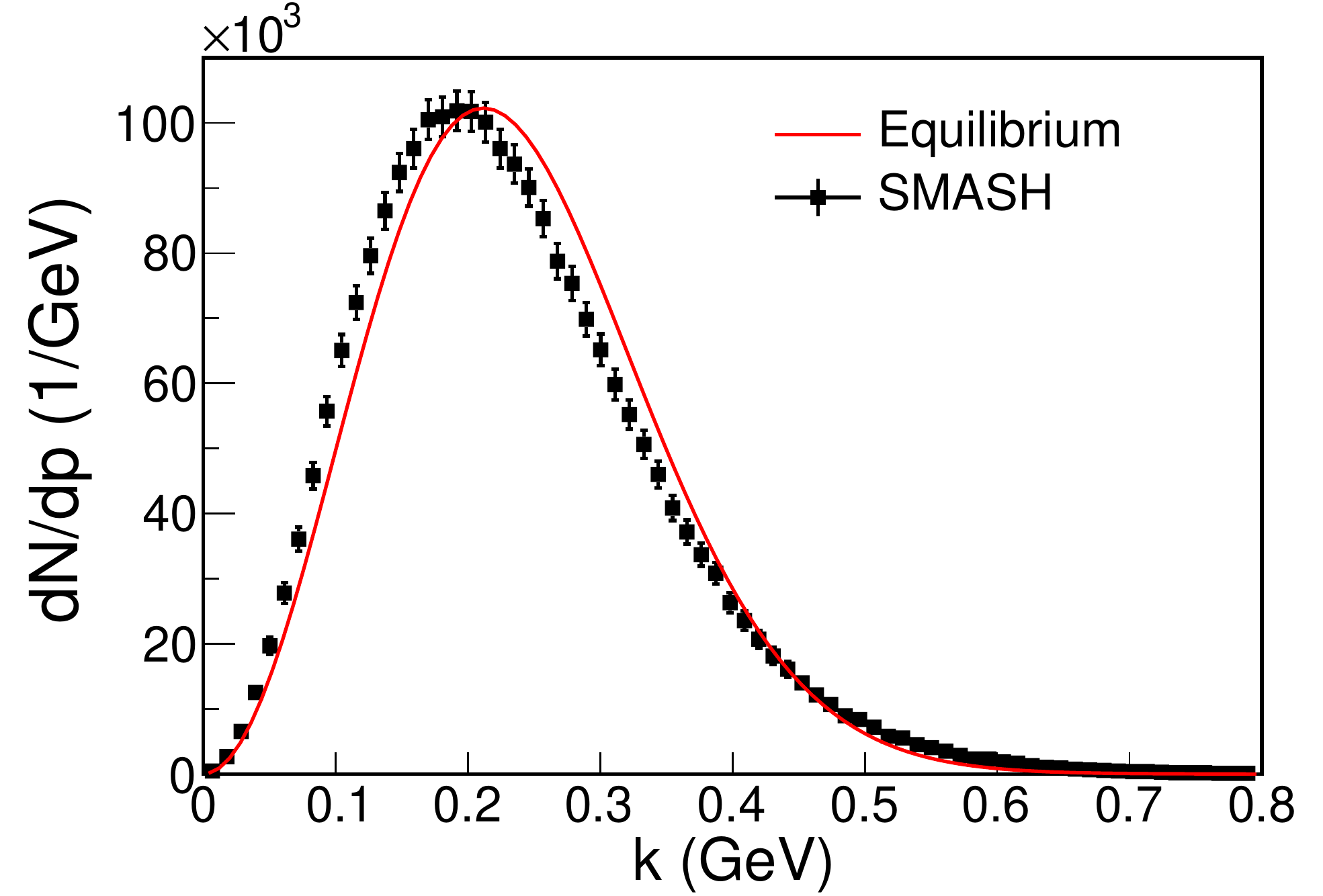}
\caption{\label{fig:distro} Distributions of particles for a relativistic massive gas at different times. We compare the outcome from SMASH and the theoretical prediction for a thermal distribution. From left to right, and
top to bottom, the times read $t=0.1,2.5,5,10,15,20$ fm, respectively.}
\end{center} 
\end{figure*}

As said, our initial state is a gas in equilibrium, where the standard Boltzmann distribution of thermodynamics can be applied. The distribution function contains two parameters: the temperature and chemical potential
(or fugacity). The latter is needed due to the absence of number-changing processes. Taking into account that collisions maintain equilibrium at early times (when $\Gamma \gg H$), it is possible to compute
the values of these two parameters by noting that the evolution of the particle density is $n (T,\mu)=n_0 (T_0,\mu_0) (a_0/a)^3$, and assuming an adiabatic evolution where the entropy per particle ($s/n$) is constant 
in the evolution. These two conditions provide a system of two coupled equations for the values of $T$ and $\mu$. The evolution of the particle density gives
\begin{equation} T  e^{\frac{\mu(T)}{T}} K_2 (m/T) = T_0  e^{\frac{\mu_0}{T_0}} K_2 (m/T_0)  \left( \frac{a_0}{a} \right)^3  \ ,  \label{eq:T}  \end{equation}
and the condition $s/n=s_0/n_0$ provides
\begin{equation}
\label{eq:mu} \frac{\mu(T)}{T}= \frac{ \mu_0}{T_0} + \frac{m}{T} \frac{K_1(m/T)}{K_2(m/T)}  - \frac{m}{T_0} \frac{K_1(m/T_0)}{K_2(m/T_0)} \ , 
\end{equation}
where standard thermodynamical relations have been used (see e.g.~\cite{letessier2002hadrons}).

It is illustrative to take the ultrarelativistic and nonrelativistic limits of these equations to recover the textbook cases,
\begin{equation} \label{eq:evolT}
 T(t) = \left\{ 
 \begin{array}{ccc}
    T_0 \frac{a_0}{a(t)}  & \textrm{for } & m \ll T \ , \\ 
    T_0 \left( \frac{a_0}{a(t)} \right)^2   & \textrm{for } & m \gg T \ ,
 \end{array}
 \right. 
\end{equation}
which tells us that in an expanding gas of ultrarelativistic (nonrelativistic) particles, the equilibrium state is maintained with a temperature scaling as $T \sim 1/a$ ($T \sim 1/a^2$).
It is possible to check that we also reproduce the evolution of the chemical potential in these two limiting cases~\cite{Bernstein}.

We check the expected equilibrium evolution using SMASH with relativistic particles with a mass of $m=2$ GeV. The cross section is assumed to be constant $\sigma =40$ mb, and the initial configuration is a Boltzmann distribution with temperature 
$T_0=0.4$ GeV and vanishing chemical potential $\mu_0=0$. The geometrical conditions are the same as in the last section. We use 20 events to reduce statistical uncertainty.

In Fig.~\ref{fig:distro} we present SMASH results at different times together with the expected equilibrium distribution. In the upper panels we check that the SMASH results compare perfectly well to the 
predicted equilibrium distribution with parameters given by Eqs.~(\ref{eq:T},\ref{eq:mu}). In the lower panels we observe that deviations occur, which increase with time. This is the effect of the decoupling, which 
roughly happens between $t=5$ fm and $t=10$ fm by looking at the panels. Using the naive condition $\Gamma \sim H$, we numerically obtain an approximate guess of $t_D \sim 8$ fm\footnote{$\Gamma$ is given 
in SMASH by counting the number of collisions per unit time.}. Although this is a very crude estimate, it is consistent with
the situation shown in Fig.~\ref{fig:distro}.

In fact, it is possible to determine the form of the distribution function after decoupling. One simply needs to take into account the unavoidable momentum redshift. For any time after decoupling $t>t_D$
\ba \label{eq:decoupling} f_{dec} (t,k) & = & f_{eq} \left( t_D, k \frac{a}{a_D} \right) \\ 
&\propto & \exp \left[ -\frac{\sqrt{(ka/a_D)^2+m^2}}{T_D} \right] \ . \nonumber \ea

Before fitting our results to this form let us consider the ultrarelativistic and the nonrelativistic limits:
\begin{equation} 
 f_{dec} (t,k) \propto \left\{ 
 \begin{array}{lll}
  \exp \left( - \frac{ka}{T_D a_D} \right) & \textrm{for } & m\ll T \\
  \exp \left[ - \frac{k^2}{2m} \frac{a^2}{T_D a^2_D} \right] & \textrm{for } & m \gg T	\ . 
  \end{array}
  \right.
\end{equation}

These functions mimic the equilibrium distributions with a temperature evolving in the same way as in Eq.~(\ref{eq:evolT}). Therefore, we recover the result that for these limiting cases, the evolution after decoupling
is indistinguishable from an equilibrium situation\footnote{In a cosmological scenario this claim is not actually true, as the effective degeneracy factor $g(T)$ of the distribution function is itself a function of temperature,
thus producing a difference between the two cases.}.
 
\begin{figure}[htp]
 \begin{center}
 \includegraphics[scale=0.36]{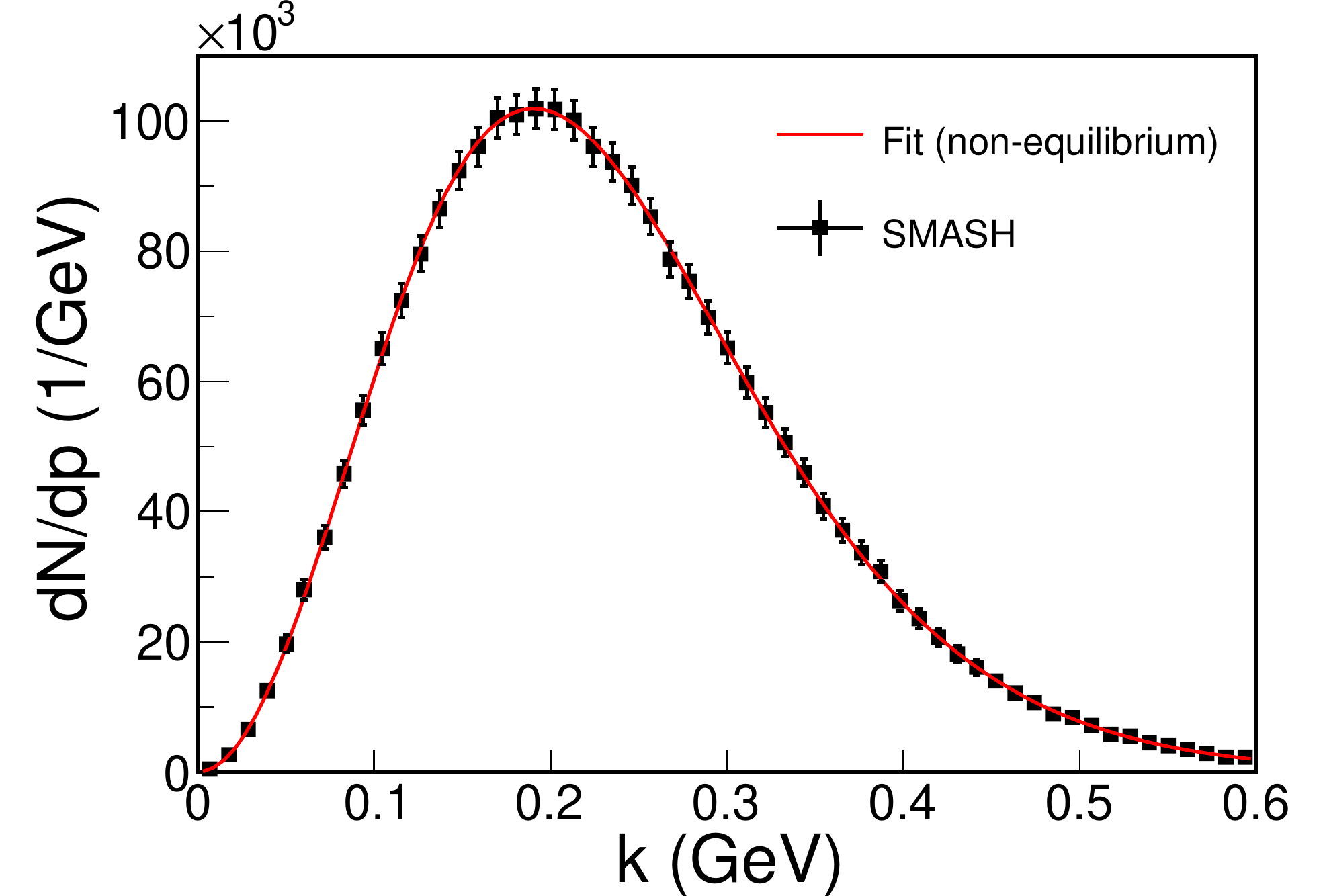}
  \caption{\label{fig:newfit} Fit of the particle distribution at $t=20$ fm to a nonequilibrium distribution of the form (\ref{eq:decoupling}).}
 \end{center} 
 \end{figure}

Let us take the distribution of our massive particles at some large time $t=20$ fm (this would correspond to the detection or measuring time). 
After checking that the distribution cannot be described by an equilibrium distribution (a $\chi^2$ test over the fit gives values around 6-7 per degree of freedom), we proceed to fit 
the nonequilibrium distribution to~(\ref{eq:decoupling}). We obtain a very good fit which is shown in Fig.~\ref{fig:newfit}.

We should stress that the fitting parameters $a_D$ and $T_D$ are not really independent, but they are related by Eq.~(\ref{eq:T}). However, as the combined implementation of such a nonlinear constraint
is involved, we have opted to assume independent parameters, and test the consistency of the results {\it a posteriori} by obtaining the freeze-out time from each of them.

From the fit we obtain a decoupling temperature of $T_D=0.32\pm 0.02$ GeV. Using Eq.~(\ref{eq:T}) and the explicit form of $a(t)=\exp(\beta t^2/2)$ we obtain a freeze-out time of $t_D=5.5^{+0.8}_{-0.9}$ fm. 
The fit also gives the value of $(a/a_D)^2=41 \pm 4$, corresponding to a freeze-out time of $t_D=5.2^{+0.8}_{-0.9}$ fm, which is quite close to the previous number, so we can provide a final average value of $t_D=5.3\pm0.6$ fm for
the freeze-out time. Notice that this time is consistent with the situation seen in Fig.~\ref{fig:distro}, although smaller 
than the crude estimate obtained by simply comparing $\Gamma$ and $H$. One has to be careful when interpreting these numbers, because all of them assume a sharp freeze-out process, which is certainly not happening 
in our case, where a smooth decoupling occurs in time\footnote{Our exponential parametrization of the scale factor helps to have a rather sharp freeze-out process. However, one cannot avoid still having a certain number of collisions 
even at times as large as $t=20$ fm.}. 

\begin{figure*}[htp]
\begin{center}
 \includegraphics[scale=0.36]{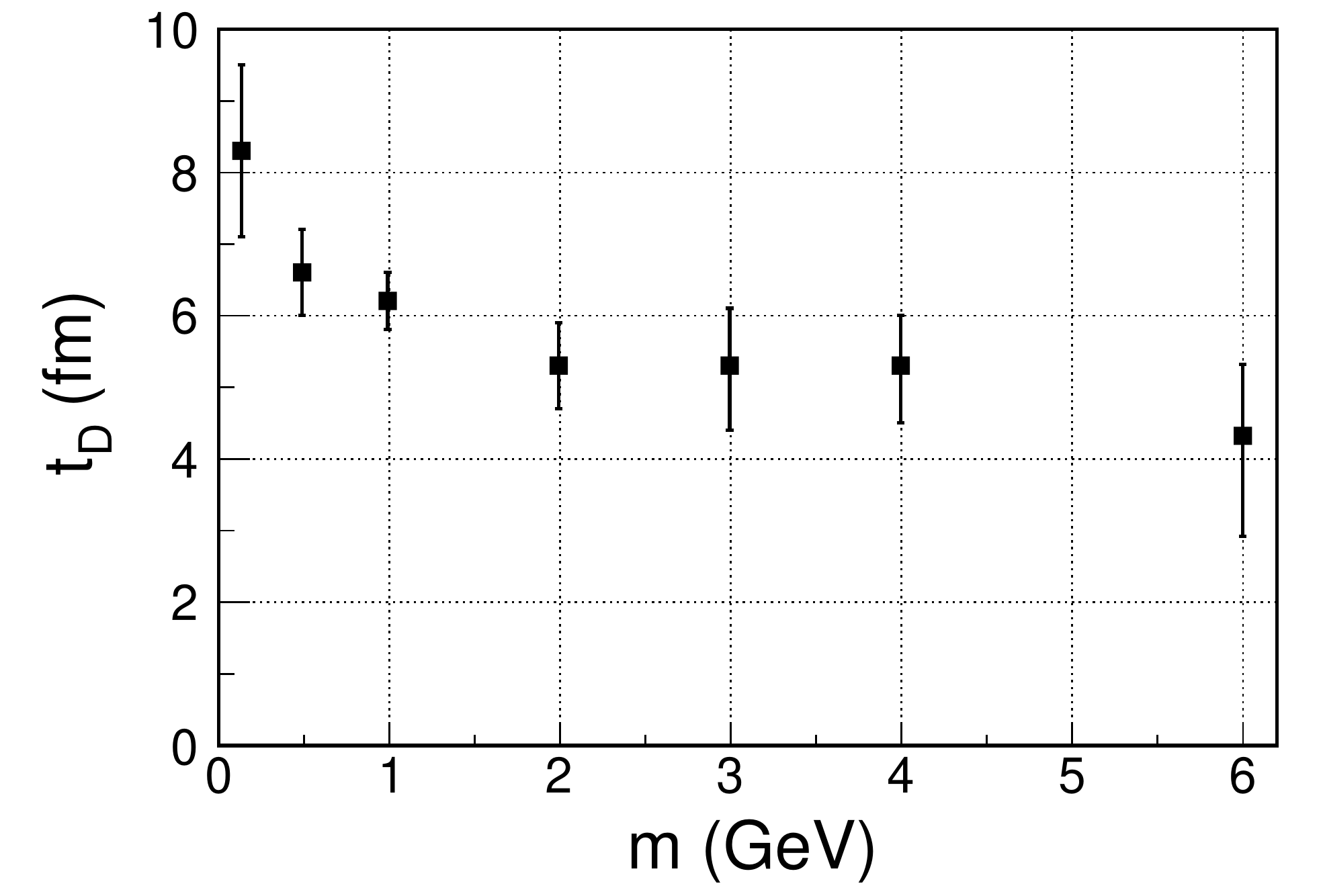}
 \includegraphics[scale=0.36]{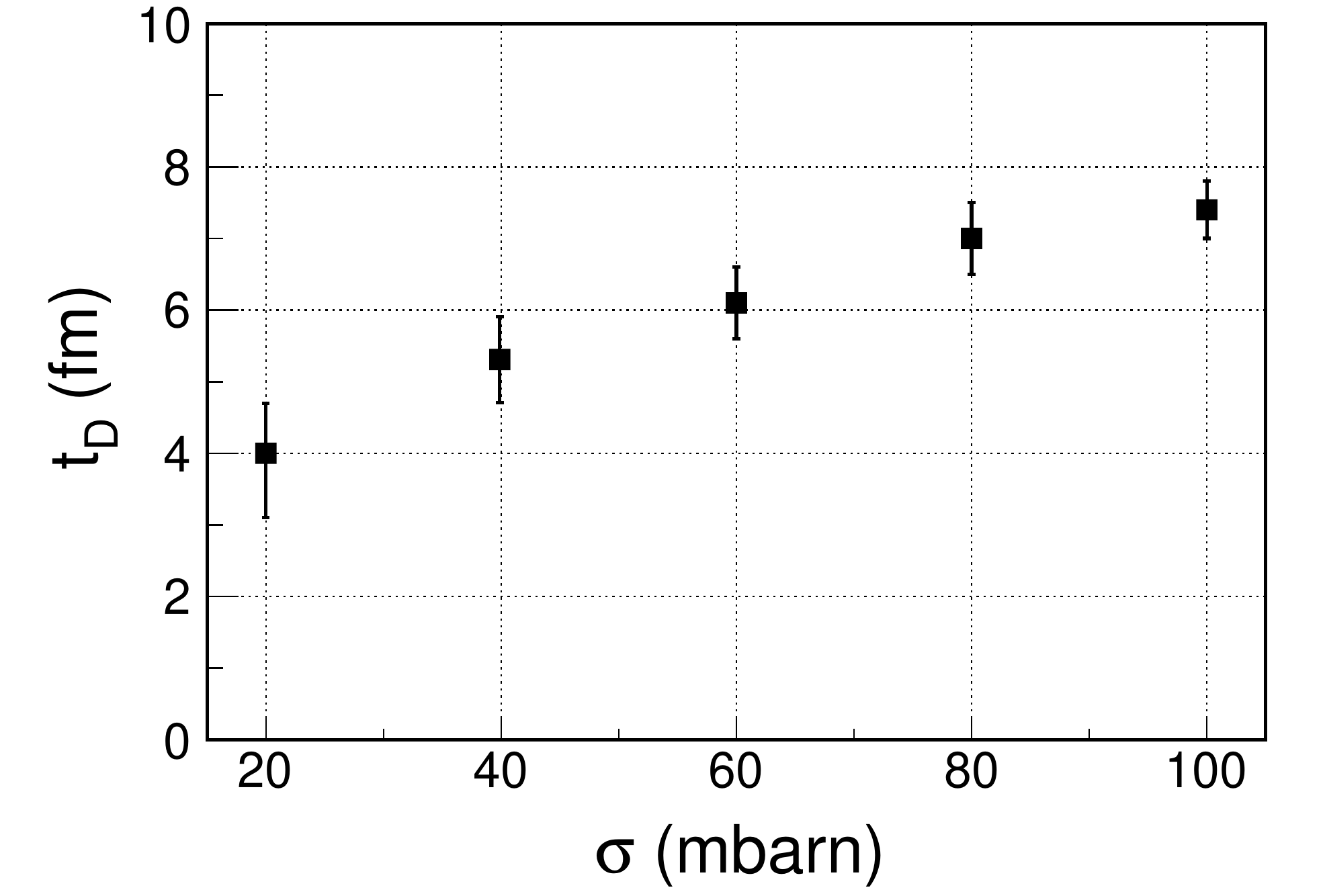}
\caption{\label{fig:masscross} Left panel: Decoupling time $t_D$ as a function of the particle mass (keeping a constant cross section of $\sigma=40$ mbarn). Right panel: Decoupling time $t_D$ as 
a function of the cross section (with a fixed mass of $m=2$ GeV).}
\end{center}
\end{figure*}

  We perform a parametric study of the freeze-out time by varying the mass of the particle and the interaction cross section. We present our results in Fig.~\ref{fig:masscross}. From this figure
we observe a slight dependence of the decoupling time on the particle mass (for smaller masses this dependence is more pronounced). This can be explained due to the fact that when the mass is increased the 
relative speed of the particles in the collision, and the particle density, are reduced. The combined effect is a decrease of $\Gamma$, which favours the appearance of the decoupling at earlier times. 
We also see a rather systematic increase of the decoupling time with the total cross section. This is also expected because for larger cross sections $\Gamma$ increases, producing 
a delay in the decoupling of particles. 

 If we had taken the condition $\Gamma=H$ to define the freeze-out times, we would have obtained systematically larger values than those shown in Fig.~\ref{fig:masscross}. A similar condition has been proposed 
in the context of RHICs trading the Hubble rate by the expansion velocity of the fireball surface~\cite{Heinz:1987ca,Lee:1987ku}. A slightly different criterion is used in Ref.~\cite{Mayer:1997mi, Hung:1997du}, 
$\xi \Gamma=H$, where $\xi$ is a parameter to be determined. Using the values of $t_D$ from Fig.~\ref{fig:masscross}, we can easily extract the corresponding $\xi$. We observe that for lower 
masses $\xi$ is closer to one, but it quickly decreases to $\xi \simeq 0.3-0.4$ for $m > 0.5$ GeV. As a function of the cross section, $\xi$ is rather stable around $\xi \simeq 0.4$, increasing up to $\xi \simeq 0.5$
for $\sigma=20$ mbarn.

 From this simple toy model for decoupling, we can already observe a rather clear dependence of the freeze-out time as a function of the particle properties. Using a more refined model including many interacting species
one might address the freeze-out features of RHICs, and verify or falsify some of the hypotheses used to describe this mechanism. In this letter we content ourselves
to present the real possibility of such a model and illustrate how the FRW expansion can provide a basis to address the freeze-out of particles in the SMASH transport approach.

%%%%%%%%%%%%%%%%%%%%%%%%%%%%%%%%%%%%%%%%%%%%%%%%%%%%%%%%%%%%%%%%%%%%%%%%%%%%%%%%%%%%%%
%%%%%%%%%%%%%%%%%%%%%%%%%%%%%%%%%%%%%%%%%%%%%%%%%%%%%%%%%%%%%%%%%%%%%%%%%%%%%%%%%%%%%%

\section{\label{sec:summary}Conclusions and Outlook}

In this letter we have reported our results on the solutions of the Boltzmann equation for a relativistic gas in a FRW spacetime evolution using the SMASH transport approach. 
For this metric there exists a particular exact solution of the Boltzmann equation which admits an analytical form~\cite{Bazow:2015dha,Bazow:2016oky}. 
We have compared the outcome of SMASH using the same initial condition and found a very good agreement for all times. Given that the initial condition is far from equilibrium, 
this agreement provides a non-trivial check for the SMASH code. We have also presented the solution of the Boltzmann equation for the ``two-mode initial condition'' introduced in Ref.~\cite{Bazow:2015dha}. 
In this case no analytical solution is known, but a numerical solution of the Boltzmann equation---written in terms of a system of coupled equations for the moments of the distribution function---is
not difficult to obtain. We also find a very good comparison between both approaches for all times.

Thanks to the versatility of SMASH we have solved the transport equation for more general cases. In particular, we have presented here the solution for massive particles, with a
Hubble expansion which produces a decoupling of particles when the interaction rate falls below the expansion rate. As long as $\Gamma \gg H$, the evolution maintains the equilibrium distribution
of particles, thanks to the collisions among them. In this limit we have seen that the predicted equilibrium distributions agree very well with the SMASH outcome. At decoupling (when $\Gamma \sim H$)
the distribution of particles departs from equilibrium due to the absence of collisions. Taking into account the momentum redshift of the particles, we have extracted the value of the freeze-out time which
is consistent with the time at which the particle distribution departs
from the equilibrium one. Finally we have presented an analysis of the freeze-out time as a function of the mass of the particle and the interaction cross section.

An interesting extension of this ``toy model'' for the freeze-out is to enlarge the particle content of the system implementing more realistic interactions according to the physics happening 
in relativistic heavy-ion collisions. This would allow access, in a more systematic manner, to the freeze-out of different species in a quantitative way and explore the hypothesis 
of a collective decoupling at fixed temperature versus a sequential freeze-out. With the results presented in this letter in a simplified scenario, we have provided indications that a dependence
of the freeze-out time on the particles' properties does exist.

%%%%%%%%%%%%%%%%%%%%%%%%%%%%%%%%%%%%%%%%%%%%%%%%%%%%%%%%%%%%%%%%%%%%%%%%%%%%%%%%%%%%%%
%%%%%%%%%%%%%%%%%%%%%%%%%%%%%%%%%%%%%%%%%%%%%%%%%%%%%%%%%%%%%%%%%%%%%%%%%%%%%%%%%%%%%%

\section{Acknowledgements} 

We acknowledge discussions with Mauricio Martinez and Jorge Noronha. We thank the DAAD and their RISE internship scheme which enabled this research to happen. 
We would also like to acknowledge funding of a Helmholtz Young Investigator Group VH-NG-822 from
the Helmholtz Association and GSI. This work was supported by the Helmholtz International Center for the Facility for Antiproton and Ion Research (HIC for FAIR) within the
framework of the Landes-Offensive zur Entwicklung Wissenschaftlich-Oekonomischer Exzellenz (LOEWE) program launched by the State of Hesse. J.M.T.-R. acknowledges support 
from Project No. FPA2013-43425-P (Spain).

\end{document}